\documentclass{desyproc}
\usepackage{psfrag,epsfig,graphicx,graphics}
\usepackage{wrapfig}
\usepackage{amssymb}
\def\stmath{}
\def\alert{}
\def\aut{}
\def\structure{}
\def\twist{}
\def\twist3{}

\def\vssp{\vspace{.15cm}}
\def\vsspEq{\vspace{.0cm}}

\def\vsssp{\vspace{.15cm}}

\def\slashchar#1{\setbox0=\hbox{$#1$}
   \dimen0=\wd0
   \setbox1=\hbox{/} \dimen1=\wd1
   \ifdim\dimen0>\dimen1
      \rlap{\hbox to \dimen0{\hfil/\hfil}}
      #1
   \else
      \rlap{\hbox to \dimen1{\hfil$#1$\hfil}}
      /
   \fi}

\newcommand{\R}{\mathbb{R}}

\newcommand{\muF}{\mu_F^2}
\newcommand{\muFa}{\mu_{F_1}^2}
\newcommand{\muFb}{\mu_{F_2}^2}

\newcommand{\rb}{\underline{r}}
\newcommand{\kb}{\underline{k}}

\def\lsim{\raise0.3ex\hbox{$<$\kern-0.75em\raise-1.1ex\hbox{$\sim$}}}
\def\gsim{\raise0.3ex\hbox{$>$\kern-0.75em\raise-1.1ex\hbox{$\sim$}}}
\def\noi{\noindent}

\def\bei{\begin{itemize}}
\def\ei{\end{itemize}}
\def\bea{\begin{eqnarray}}
\def\beqa{\begin{eqnarray}}
\def\eea{\end{eqnarray}}
\def\eqa{\end{eqnarray}}
\def\beas{\begin{eqnarray*}}
\def\eeas{\end{eqnarray*}}
\def\beqas{\begin{eqnarray*}}
\def\eqas{\end{eqnarray*}}
\def\beq{\begin{equation}}
\def\eq{\end{equation}}
\def\eeq{\end{equation}}
\def\beqd{\begin{displaymath}}
\def\eeqd{\end{displaymath}}
\def\eqd{\end{displaymath}}

\def\beeq{\begin{eqnarray}} \def\eeeq{\end{eqnarray}}

\begin{document}
%------------------------------------
\title{Hard exclusive processes: theoretical status}

%for single authors the superscripts are optional
\author{{\slshape Samuel Wallon}\\[1ex]
LPT, Universit{\'e} Paris-Sud, CNRS, 91405, Orsay, France
\ {\em \&}\\
UPMC Univ. Paris 06, facult\'e de physique, 4 place Jussieu, 75252 Paris Cedex 05, France}

% if the proceedings are available online (e.g. at Indico)
% please enter the contribution ID or file_name below for the DOI
\contribID{32}
%\contribID{smith\_joe}

% TO THE CONFERENCE EDITORS: 
% please update the following information      
% before sending the template to the authors
\confID{1407}  % if the conference is on Indico uncomment this line
\desyproc{DESY-PROC-2009-03}
\acronym{PHOTON09} % if you want the Acronym in the page footer uncomment this line
\doi  % if there is an online version we will register DOIs

\maketitle

\begin{abstract}
We review  the descriptions of hard exclusive processes
based on QCD factorization. 
\end{abstract}

\section{Introduction}

Since a decade, there have been much developments in hard exclusive processes, based on collinear factorization.
This was initiated by
    form factors studies more than 30 years ago, leading to %the introduction of 
the concept of Distribution Amplitudes (DA) \cite{Chernyak:1977fk}, which describes the partonic content of a hadron facing an elastic scattering off a hard photonic probe.
%, and their QCD evolution \cite{ERBL, Chernyak:1977fk}. 
These  DAs %, corresponding to the production of a single hadron, 
 were then extended to Generalized Distribution Amplitudes (GDA) \cite{Watanabe:1982ue, Muller:1994fv, GDA} in which two or more  hadrons 
 are produced.
% (this still fixed number of hadrons is two or three in practice \cite{PireTer}.
%In a somehow independent way, 
Independently, starting from inclusive DIS 
%and its partonic intepretation, 
%based  on the optical theorem 
which relates Parton Distribution Functions (PDFs) to the discontinuity of the forward 
$\gamma^* p \to \gamma^* p$  amplitude, it was shown \cite{MullerGPD, Muller:1994fv} that the partonic interpretation remains valid for the Deep Virtual Compton Scattering (DVCS) amplitude $\gamma^* p \to \gamma \, p$ itself, leading to the concept of 
Generalized Parton Distributions (GPDs), and to more
 general exclusive processes studies.
  In a parallel way, tremendous progresses in experimental facilities
(high luminosity beams, improvement of detectors...) opened the way to studies and measures with increasing precision of these non forward matrix  elements \cite{GPDexp, expRhoLow, rho0HERMES, expVectorHigh}.

\section{The illuminating example of $\rho-$electroproduction}

%\subsection
\noi{\it DVCS and GPD} \cite{reviewGPD} 
\vsssp

\psfrag{H}[cc][cc]{\hspace{-.1cm} \vspace{.3cm} $ H$} 
\psfrag{S}[cc][cc]{$\hspace{.1cm} S$} 
\psfrag{gas}[rc][lc]{$\gamma^*(q)$}
\psfrag{g}[cc][cc]{$\gamma$}
%\psfrag{D}[cc][cc]{$\Delta$}
\psfrag{D}[cc][cc]{}
\psfrag{pp}[rt][rt]{$p=p_2-\Delta$}
\psfrag{ppp}[lt][lt]{$\,p'=p_2+\Delta$}
%%%%%%%%I use conventions opposite to Markus%%%%%%%%%
%
%%%%down along -   %%%%%%%%%%%%
\psfrag{pl}[rc][Br]{$\int d^4k \qquad k$}
\psfrag{pr}[cc][Bc]{$\ k+\Delta$}
%%%%above along +  %%%%%%%%%%%%
\psfrag{pmq}[cc][cc]{$+ \, -$}
\psfrag{mu}[cc][cc]{$+$}
\psfrag{pmq}[cc][cc]{}
\psfrag{mu}[cc][cc]{}
%
%
%\vspace{-.3cm}
\begin{figure}[htb]
\scalebox{1}{
\begin{tabular}{ccc}
\hspace{1.5cm}\raisebox{-.44 \totalheight}{\includegraphics[height=5cm]{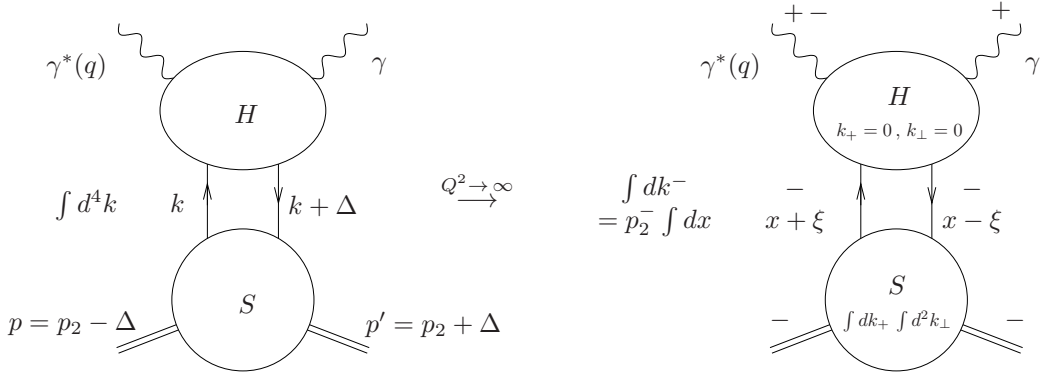}} 
& \hspace{-.1cm} $\stackrel{Q^2 \to \,\infty}{\longrightarrow}$ \hspace{.5cm}&
%\psfrag{H}[cc][cc]{\scalebox{.9}{$\begin{array}{c}\alert{H}\\ {\tiny k_+=0\, , k_\perp=0} \end{array}$}} 
\psfrag{H}[cc][cc]{\scalebox{1}{\begin{tabular}{c}$\alert{H}$\\ \scalebox{.7}{$k_+=0\, , \,k_\perp=0$} \end{tabular}}}
%\psfrag{S}[rc][lc]{$\begin{array}{c}{\stmath S}\\ \int k_+\, , \int d^2 k^2_\perp \end{array}$} 
\psfrag{S}[cc][cc]{\scalebox{1}{\begin{tabular}{c}${\stmath S}$\\ \scalebox{.7}{$\int d k_+\,  \int d^2 k_\perp$} \end{tabular}}}
%
%%%%down along -   %%%%%%%%%%%%
\psfrag{pl}[rc][Br]{$ \begin{array}{c} \int dk^-\\  =p_2^- \int dx \end{array} \quad \begin{array}{c}-\\  x+\xi \end{array}$}
\psfrag{pr}[cc][Bc]{$\begin{array}{c}-\\  x-\xi \end{array}$}
\psfrag{pp}[rc][rc]{$-$}
\psfrag{ppp}[cc][lc]{$-$}
%%%%above along +  %%%%%%%%%%%%
\psfrag{pmq}[cc][cc]{$+ \, -$}
\psfrag{mu}[cc][cc]{$+$}
\hspace{2.6cm}
\raisebox{-.44 \totalheight}{\includegraphics[height=5cm]{wallon_samuel.factGPD_HsD.eps}} 
\end{tabular}
}
\caption{Factorization of the DVCS amplitude in the hard regime. The signs $+$ and $-$ indicate corresponding flows of large momentum components.}
\label{Fig:factGPD}
\end{figure}
%%%%%%%%%%%%%%%%%%%%%%%%%%%%%%%%%%%%%%%%%%%%%%%%%%
%%%%%%%%%%%%%%%%%%%%%%%%%%%%%%%%%%%%%%%%%%%%%%%%%%
The factorization of the DVCS amplitude in the large $Q^2$ limit follows  two steps.
First, one should factorize it in momentum space. This can be set up more easily when 
using the Sudakov decomposition (introducing two light-cone vectors $p_{1(2)}$ ($+(-)$ directions)
%and $p_2$ ($-$ direction)
 with $2 \,p_1 \cdot p_2=s$)
\beq
\label{Sudakov}
\begin{array}{ccccccc}
k &=& \alpha \, p_1 &+& \beta \, p_2 &+& k_\perp 
\\
& & + & & - & & \perp
\end{array}
\eq
In the limit $Q^2 \to \infty$, the only component of the momentum $k$ to be kept in the hard blob $H$ is $k_-\,.$ In particular, this means that the quark-antiquark pair entering $H$
 can be considered as being collinear, flying in the direction of the 
$p_2$ momentum. Therefore,  the amplitude reads
\beqas
\begin{array}{ccccc}
\hspace{0cm}\int d^4 k \,\, S(k,\, k+\Delta) \, H(q,\, k,\, k+\Delta) &\!\!=\!\!&\! \int dk^-&\hspace{-.2cm}\stmath{\int dk^+ d^2 k_\perp \, S(k,\, k+\Delta)} &\hspace{-.2cm} \alert{H(q, \,k^-,\, k^- +\Delta^-)} \,,
\end{array}
\eqas
as illustrated in Fig.\ref{Fig:factGPD}.
% with 
% $p_2= \frac{1}{2}(p+p') \,, \ \Delta = p'-p$
%
%
The  Fierz identity in spinor and color space then shows that the DVCS amplitude completely factorizes, and reads symbolically:
$%$
 {\cal M} = {\stmath{\rm GPD}} \otimes \alert{\mbox{Hard part}}\,.
$%$
%M\" uller et al. '91 - '94; Radyushkin '96; Ji '97

\noi{\it $\rho-$meson production: from the wave function to the DA}
\vsssp

\psfrag{gas}[cc][lc]{$\!\!\!\!\gamma^*(q)$}
\psfrag{M}[cc][lc]{$\,\,\,\alert{M}$}
\psfrag{P}[cc][lc]{$\,\,\,\stmath{\Psi}$}
\psfrag{pp}[cB][lc]{$p$}
\psfrag{ppp}[lB][lc]{$\,\,\,\,\, \, p'$}
\psfrag{r}[ct][ct]{$\,  \stmath{\rho}$}
\psfrag{V}[cB][cc]{$p_\rho$}
%%%%above along +  %%%%%%%%%%%%
\psfrag{pmq}[cc][cc]{}
\psfrag{mu}[Bc][lc]{$\hspace{-.4cm}\int d^4 \ell \quad \ell$}
\psfrag{md}[Bc][cc]{$\ell-p_\rho$}
%%%%%%%
\begin{figure}[htb]
\vspace{-.3cm}
\begin{tabular}{ccc}
\hspace{1cm}\raisebox{-.44 \totalheight}{\includegraphics[height=2.8cm]{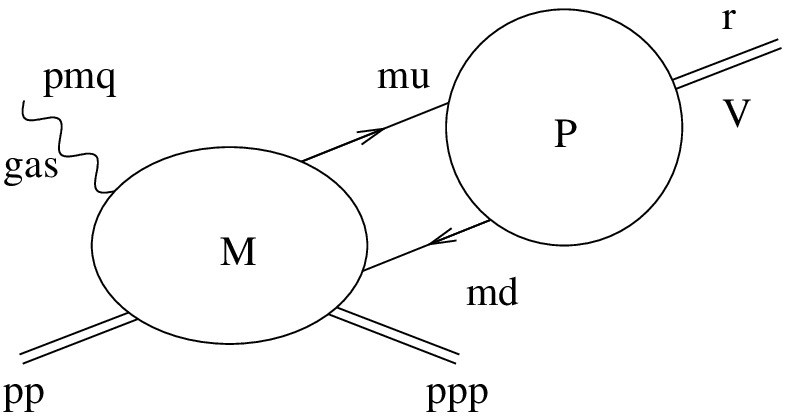}} 
& \hspace{-.5cm}  \raisebox{-.44 \totalheight}{$\stackrel{Q^2 \to \, \infty}{\longrightarrow}$} \hspace{-.1cm} &
\psfrag{M}[cc][cc]{\scalebox{1}{\begin{tabular}{c}$\alert{M}$\\ \scalebox{.7}{$\!\!\ell_-=0\, , \,\ell_\perp=0$} \end{tabular}}}
\psfrag{P}[cc][cc]{\scalebox{1}{\begin{tabular}{c}${\stmath \Psi}$\\ \scalebox{.7}{$\int d \ell_-\,  \int d^2 \ell_\perp$} \end{tabular}}}
\psfrag{gas}[cc][lc]{$+ \, -$}
%%%%above along +  %%%%%%%%%%%%
\psfrag{pmq}[cc][cc]{}
%\psfrag{mu}[Bc][lc]{\raisebox{.44 \totalheight}{$ \begin{array}{c}\footnotesize{\begin{array}{c} \int d\ell^+\\  =p_1^+ \int dz \end{array}}\\ \\ \quad z \, \, +\end{array}$}}
%
\psfrag{mu}[Bc][lc]{\hspace{-1.2cm}\raisebox{.35 \totalheight}{$ \begin{array}{cc}\footnotesize{\begin{array}{c} \int d\ell^+ \\  =p_1^+ \int du \end{array}}&  \vspace{-.2cm} \hspace{-.2cm} u \, \, +\end{array}$}}
%\begin{array}{c}+\\  z \end{array}}$}
\psfrag{md}[Bc][cc]{$\!-\bar{u} \,\,\,+$}
\psfrag{r}[ct][ct]{$\,  \stmath{\rho}$}
\psfrag{V}[cB][cc]{$+$}
%
%%%%below along -  %%%%%%%%%%%%
%
\psfrag{pp}[cB][rc]{$-$}
\psfrag{ppp}[cB][lc]{$-$}
\raisebox{-.44 \totalheight}{\includegraphics[height=3cm]{wallon_samuel.factM_DA.eps}}
\end{tabular}
\vspace{-.1cm}
\caption{Factorization of the amplitude of hard $\rho-$electroproduction.}
\label{Fig:FactRho}
\end{figure}
We now replace the produced photon by a $\rho-$meson,  described in QCD by its \alert{wave function} $\stmath\Psi$ which reduces in \structure{hard processes} to its \alert{Distribution Amplitude}. As 
for DVCS, in the limit $Q^2 \to \infty\,,$ the amplitude of diffractive
electroproduction of a $\rho-$meson can be written as  
\beq
\label{DAfact}
\hspace{0cm}\int d^4 \ell \,\,  M(q,\, \ell ,\, \ell -p_\rho) \,
 \Psi(\ell,\, \ell-p_\rho) \,= \int d\ell^+  \alert{M(q, \,\ell^+,\, \ell^+ -p_\rho^+)}
\stmath{\int d\ell^-\hspace{-.3cm} \int\limits^{\scalebox{.6}{$|\ell_\perp^2| < \muF$}}
d^2 \ell_\perp \, \Psi(\ell,\, \ell-p_\rho)} 
\eq
 (see Fig.\ref{Fig:FactRho}). This factorization involves the $\rho-$wave function integrated over $\ell_\perp$ (and $\ell^-$), which is the DA already
involved in the partonic description of the hard meson form factor \cite{Chernyak:1977fk}.
\vssp

%(see
%{\aut \small %Brodsky, Farrar '73; 
% Chernyak, Zhitnitsky '77; Brodsky, Lepage '79; Efremov, Radyushkin '80; ...} in the case of form-factors studies)

\noi{\it $\rho-$meson production: factorization with a GPD and a  DA} \cite{fact}
\vsssp

\psfrag{M}[cc][lc]{$\,\,\,\stmath{\Psi}$}
\psfrag{S}[cc][lc]{$\stmath{S}$}
\psfrag{H}[cc][lc]{$\alert{H}$}
\psfrag{gas}[cc][lc]{$\gamma^*(q)$}
\psfrag{g}[cc][cc]{$\gamma$}
\psfrag{D}[cc][cc]{}
\psfrag{pp}[cc][cc]{$p$}
\psfrag{ppp}[cc][cc]{$p'$}
\psfrag{r}[cc][cc]{}
\psfrag{V}[cc][cc]{$p_\rho$}
\psfrag{gas}[rc][lc]{$\gamma^*(q)$}
\psfrag{g}[cc][cc]{$\gamma$}
\psfrag{D}[cc][cc]{}
\psfrag{pp}[rt][rt]{\footnotesize{$p=p_2-\Delta$}}
\psfrag{ppp}[lt][lt]{\footnotesize{$p'=p_2+\Delta$}}
%
%
%%%%%%%%I use conventions opposite to Markus%%%%%%%%%
%
%
%%%%above along +  %%%%%%%%%%%%
\psfrag{pmq}[cc][cc]{}
\psfrag{mu}[Bc][lc]{$\hspace{-.4cm}\int d^4 \ell \quad \ell$}
\psfrag{md}[Bc][cc]{\raisebox{-.1cm}{$\ell-p_\rho$}}
%
%%%%down along -   %%%%%%%%%%%%
\psfrag{pl}[rc][Br]{$\int d^4k \qquad k$}
\psfrag{pr}[cc][Bc]{$\hspace{.4cm} k+\Delta$}
\begin{figure}[htb]
\begin{tabular}{ccc}
\scalebox{1}{\hspace{1cm}\raisebox{-.44 \totalheight}{\includegraphics[height=4.5cm]{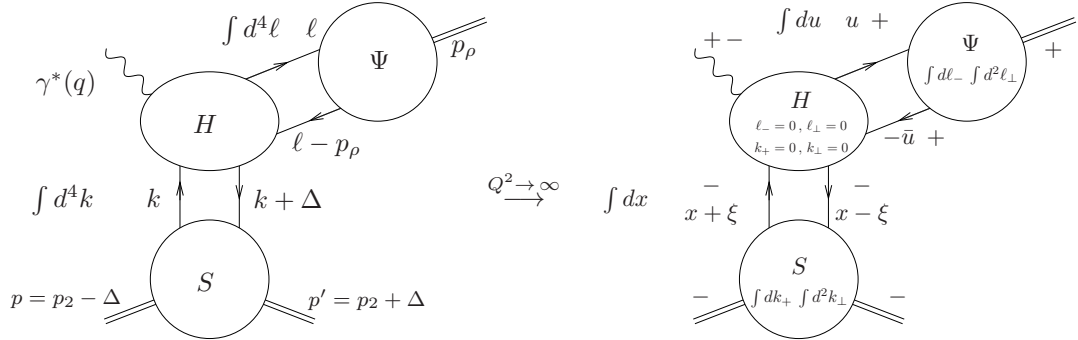}}}
& \hspace{-.6cm} \raisebox{-.44 \totalheight}{$\stackrel{Q^2 \to \, \infty}{\longrightarrow}$}  &
\psfrag{H}[cc][cc]{\raisebox{.2cm}
{\scalebox{1}{\begin{tabular}{c}$\alert{H}$\vspace{-.1cm}\\ \vspace{-.1cm}\scalebox{.58}{$\ell_-=0\, , \,\ell_\perp=0$} \\
\vspace{0cm}\scalebox{.58}{$k_+=0\, , \,k_\perp=0$}
\end{tabular}}}}
\psfrag{M}[cc][cc]{\scalebox{1}{\begin{tabular}{c}${\stmath \Psi}$\\ \scalebox{.7}{$\int d \ell_-\,  \int d^2 \ell_\perp$}\end{tabular}}}
\psfrag{S}[cc][cc]{\scalebox{1}{\begin{tabular}{c}${\stmath S}$\\ \scalebox{.7}{$\int d k_+\,  \int d^2 k_\perp$} \end{tabular}}}
\psfrag{gas}[cc][lc]{}
%%%%down along -   %%%%%%%%%%%%
%\psfrag{pl}[rc][Br]{$ \begin{array}{c} \int dk^-\\  =p_2^- \int dx \end{array} \quad \begin{array}{c}-\\  x+\xi \end{array}$}
\psfrag{pl}[rc][Br]{$\int dx \quad \begin{array}{c}-\\  x+\xi \end{array}$}
\psfrag{pr}[cc][Bc]{$\begin{array}{c}-\\  x-\xi \end{array}$}
\psfrag{pp}[rc][rc]{$-$}
\psfrag{ppp}[cc][lc]{$-$}
%
%
%%%%above along +  %%%%%%%%%%%%
\psfrag{pmq}[cc][cc]{$+ \, -$}
\psfrag{mu}[Bc][lc]{\hspace{-1.2cm}\raisebox{.35 \totalheight}{$ \begin{array}{cc} \int du &  u \ \, +\end{array}$}}
\psfrag{md}[Bc][cc]{$\!-\bar{u} \ \,+$}
\psfrag{V}[cc][cc]{$+$}
\hspace{1.1cm}
\scalebox{.9}{\raisebox{-.44 \totalheight}{\includegraphics[height=5cm]{wallon_samuel.factGPD_H_DAsD.eps}}}
\end{tabular}
\caption{Full factorization of the amplitude of hard electroproduction of a $\rho-$meson.}
\label{Fig:FactRhoGPD}
\end{figure}
%\begin{frame}{$\rho-$electroproduction}{
Combining the previous factorizations, one can describe the hard electroproduction
of a $\rho-$meson in a fully factorized form involving a GPD and a DA, as illustrated in Fig.\ref{Fig:FactRhoGPD}. It reads
\beqa
\label{FactGPD_DA}
%\begin{array}{ccccccc}
&& \int d^4 k \, d^4 \ell\,\,  S(k,\, k+\Delta) \, H(q,\, k,\, k+\Delta) \, \Psi(\ell,\, \ell-p_\rho)=\int dk^- d\ell^+\\
\nonumber\\
%&\!\!=\!\!&\! \int dk^- \int d\ell^+&\hspace{-.4cm}\stmath{\int dk^+ d^2 k_\perp \, S(k,\, k+\Delta)} &\hspace{-.2cm} \alert{H(q; \,k^-,\, k^- +\Delta^-; \ell^+,\, \ell^+-p_\rho^+)} & \Psi(\ell,\, \ell^-p_\rho)\\
%\frac{s}{2}\int dx \int du
 &&\hspace{-.6cm}\times\stmath{\int dk^+\hspace{0cm} \int\limits^{\scalebox{.6}{$|k_\perp^2| < \muFb$}} \! d^2 k_\perp \, S(k, k+\Delta)}  \ \ \alert{ H(q; k^-\!, k^- \!+\Delta^-; \ell^+\!, \ell^+\!-p_\rho^+)} \ \ \stmath{\int d\ell^- \hspace{-.2cm} \int\limits^{\scalebox{.6}{$|\ell_\perp^2| < \muFa$}}\!
d^2 \ell_\perp \Psi(\ell,\, \ell-p_\rho)}\,.\nonumber\\
\nonumber\\
&&\hspace{.7cm}\stmath{\mbox{GPD } F(x, \, \xi, t,\muFb)} \hspace{.9cm} \alert{\mbox{Hard part }T(x/\xi,u,\muFa,\muFb)}\hspace{.8cm} \stmath{\mbox{DA }\Phi(u,\muFa)}\,\nonumber
\eqa
\vsspEq

\noi{\it Chiral-even DA}
\vsssp

As discussed above, DAs are obtained from wave functions through   $\int d\ell^- \int d^2\ell_\perp$ integration, and thus related to \alert{non-local} correlators between fields separated by a \alert{light-like} distance $\alert{z}$ (along $p_2$, conjugated to the \alert{+} direction by {\aut Fourier} transformation). The vector correlator reads
\beqas
%\label{defDAChiralEvenV}
&&\hspace{-1cm}\langle 0|\bar u(\alert{z}) \gamma_{\mu} d(-\alert{z})|\rho^-(P,\lambda)\rangle 
 = f_{\rho} m_{\rho} \left[ p_{\mu}
\frac{e^{(\lambda)}\cdot z}{p \cdot z}
\int_{0}^{1} \!du\, e^{i (u-\bar{u}) p \cdot z} \alert{\phi_{\parallel}(u, \muF)} \right. 
\nonumber\\
&+& \left.e^{(\lambda)}_{\perp \mu}
\int_{0}^{1} \!du\, e^{i (u-\bar{u}) p \cdot z} {\stmath g_{\perp}^{(v)}(u, \muF)} 
- \frac{1}{2}z_{\mu}
\frac{e^{(\lambda)}\cdot z }{(p \cdot z)^{2}} m_{\rho}^{2}
\int_{0}^{1} \!du\, e^{i (u-\bar{u}) p \cdot z} {\aut g_{3}(u, \muF)}
\right]
\eqas
where $\phi_{\parallel}, \,g_{\perp}^{(v)},\, g_{3}$ are DAs respectively of twist 2, 3 and 4, with $p=p_1,  \, P=p_\rho\,.$ Correspondingly, the axial correlator calls for the introduction of a twist 3 DA, as
\[
%\label{defDAChiralEvenA}
\langle 0|\bar u(\alert{z}) \gamma_{\mu} \gamma_{5} 
d(-\alert{z})|\rho^-(P,\lambda)\rangle= \frac{1}{2}\left[f_{\rho} - f_{\rho}^{T}
\frac{m_{u} + m_{d}}{m_{\rho}}\right]
m_{\rho} \, \epsilon_{\mu}^{\phantom{\mu}\nu \alpha \beta}
e^{(\lambda)}_{\perp \nu} \, p_{\alpha} \, z_{\beta}
\int_{0}^{1} \!du\, e^{i \xi p \cdot z} {\stmath g^{(a)}_{\perp}(u, \muF)}\,.%,
\]
\vssp

\noi{\it Selection rules and factorization status}
\vsssp

Since for massless particle 
 chirality = + (resp. -) helicity for a (anti)particule 
and based on the fact that \structure{QED and QCD vertices are chiral even} (no chirality flip during the interaction), one deduces that 
the total helicity of a $q \bar{q}$ pair produced by a $\gamma^*$ should be 0.
Therefore, the helicity of the $\gamma^*$ equals $L^{q \bar{q}}_z$ ($z$ projection of the 
$q \bar{q}$ angular momentum).
In the pure collinear limit (i.e. twist 2), $L^{q \bar{q}}_z=\,0$, and thus
the $\gamma^*$ is longitudinally polarized.
Additionaly, 
at $t=0$ there is no source of orbital momentum from the proton coupling, which implies
that 
the helicity of the meson and of the photon should be identical.
In the collinear factorization approach, the extension to $t \neq 0$ changes nothing from the hard side, 
the only dependence with respect to $t$ being encoded in the non-perturbative correlator which defines the GPDs. This implies that 
  the above selection rule remains true. Thus, 
only   2 transitions are possible (this is called $s-$channel helicity conservation (SCHC)):
 $\gamma^*_L \to \rho_L$, for which QCD factorization \alert{ holds at t=2} at any order 
(i.e. LL, NLL, etc...) \cite{fact} and 
%{\aut \small Collins, Frankfurt, Strikman '97}
$\gamma^*_T \to \rho_T$, corresponding to twist $t=3$ at the amplitude level, for which  QCD factorization is not proven,  an explicit computation  \cite{Mank} at leading order showing in fact that the hard part has end-point singularities like
%{\aut \small Mankiewicz-Piller '00}
$\int\limits_0^1 du/u\,.$ %or $\int\limits_0^1 du/(1-u)\,.$
\vssp

\noi{\it Some solutions to factorization breaking?}
\vsssp

In order to extend the factorization theorem at higher twist, several solutions have been 
discussed. First, one may  add contributions of \alert{3-parton DAs} \cite{BB} for $\rho_T$ \cite{Anikin:2002wg, Anikin:2002uv} (of dominant twist equal 3 for $\rho_T$). 
This in fact does not solve the problem, while reducing the level of
divergency, but is needed for consistency. Next, it was suggested to keep a 
 transverse \alert{$\ell_\perp$} dependency in the $q$, $\bar{q}$ momenta, used to regulate end-point singularities, leading to
the Improved Collinear Approximation (ICA).
Soft and collinear gluon exchange between the valence quarks are responsible for large double-logarithmic effects which exponentiate. 
This is made easier when using the impact parameter space $b_\perp$ conjugated to $\ell_\perp\,,$ leading to 
{\aut Sudakov} factor
%\beq
%\label{SudakovFact}
\[
\exp [-S(u, b, Q)   ]\,.
%\eq
\]
$S$ diverges when ${\stmath b_\perp \sim O(1/\Lambda_{QCD})}$ (large transverse separation, i.e. \structure{small transverse momenta}) or %\structure{small fraction} 
${\stmath u \sim O(\Lambda_{QCD}/Q)}$ \cite{Botts:1989kf}.
%{\aut \small  Botts, Sterman '89}
This  regularizes end-point singularities for $\pi \to \pi \gamma^*$ and $\gamma \gamma^* \to \pi^0$ form factors \cite{Li:1992nu}.
%, based on the factorization approach
%{\aut \small  Li, Sterman '92}
This perturbative resummation tail effect combined 
 with an ad-hoc non-perturbative gaussian ansatz for the DAs
%\beq
%\label{GaussianDA}
\[ 
\exp [ -a^2 \, |k_\perp^2|/(u \bar{u}) ] \,,
%\eq
\]  
which gives back the usual asymptotic DA $6 \, u \, \bar{u}$ when integrating over $k_\perp$,
provides 
 practical tools for the phenomenology of meson electroproduction \cite{GK}.  
%{\aut  \small Goloskokov, Kroll '05}
\vssp

\noi{\it Chiral-odd sector}
\vsssp

%\vskip.1in

The $\pm$ {\alert{chiralities}} are defined by the decomposition
\[
%\beq
%\label{defChirality}
q_\pm(z) \equiv \frac{1}{2}(1\pm \gamma^5)q(z)\quad {\rm with}\;\;\;\;\;\;q(z)= q_+(z) + q_-(z)\,,
%\eq
\]
implying that 
{$\bar q_{\pm}(z) \gamma^\mu q_\pm(-z)$ or  $\bar q_\pm(z) \gamma^\mu \gamma^5q_\pm(-z)$} \structure{conserve chirality} (chiral-even)
while
$\bar q_{\pm}(z)\cdot 1\cdot q_\mp(-z),\,
\bar q_{\pm}(z)\cdot \gamma^5\cdot q_\mp(-z)$ or  $\bar q_\pm(z) [\gamma^\mu ,\gamma^\nu] q_\mp(-z)$ change chirality
(chiral-\alert{odd}).
In the specific case of $\rho$, the chiral odd  sector involves DAs of twists 2 and 4
for $\rho_T$ and DAs of twist 3 for $\rho_L\,.$ Correspondingly, chiral-odd 3-partons DAs 
are of dominant twist equal to 3 for $\rho_L$  \cite{BB}. 

Since QED and QCD are chiral even, chiral-odd objects can only appear in pairs. While the amplitude of $\rho_T$
electroproduction  on linearly polarized $N$ vanishes at leading twist 2 (a single gluon exchange between hard lines is not enough to prevent the vanishing of Dirac traces) \cite{DGP}, this vanishing can be avoided \cite{IPST}, for example in the electroproduction of  a $\pi^+$ and $\rho^0_T$ pair on a nucleon $N$ \cite{rhoTMounir}, the hard scale being provided by the $p_T$ of the produced mesons.

\section{Generic results  for DAs}

\noi{\it Gauge invariance}
\vsssp

The non-local correlators $\langle 0|\bar \Psi(z) \gamma_{\mu} \Psi(-z)|\rho\rangle$ are gauge invariant since they
should be understood as $\langle 0|\bar \Psi(z) \gamma_{\mu} \alert{[z,\, -z\,]}\Psi(-z)|\rho\rangle$
where $[,]$ is a {\aut Wilson} line along $p_2\,.$ This implies that even at twist 2, gluons are there, although hidden. 
The {\aut Taylor} expansion with respect to $z$ involves the covariant derivative $\stackrel{\leftrightarrow} {D_{\mu}}\,.$
This can be used for studying hard electroproduction of \alert{exotic} (non $q\bar{q}$ quantum numbers) \alert{hybrids mesons} $| q \bar{q} g \rangle$ with ${\stmath J^{PC}=1^{-+}}$ 
which \alert{cannot} be described by the quark model.
Thus, $\gamma^* p \to H^0 p$ is \alert{not} suppressed: it is \alert{twist 2}. The expected order of magnitude of the cross-section is comparable with $\rho$-electroproduction \cite{APSTWElectro}, with possible tests
%{\aut \small Anikin, Pire, Szymanowski, Teryaev, S.W. \, '04, '05}
 at {\aut JLab and Compass}.   The same conclusion applies for the process $\gamma \gamma^* \to H^0$
with the advantage of avoiding the mixing with GPDs \cite{APSTWGamma}. Tests should be
possible
% \ {\aut \small  ibid. '06}
 at {\aut BaBar, BELLE, Bepc-II}.
%\hfill {n.b.: $H^0 \to \pi \eta$}
A possible candidate for the neutral hybrid $H^0$ could be the $\pi_1(1400)$
candidate.
\vssp

\noi{\it Equations of motion}
%\centering{Equation of motion}
\vsssp

 {\aut Dirac} equation leads to
%\beq
%\label{Dirac}
$
{\stmath \langle i(\stackrel{\rightarrow}
{\slashchar{D}}(0) \psi(0))_\alpha\, \bar \psi_\beta(z)\rangle = 0}$
%\, \qquad {\scriptsize(i\stackrel{\rightarrow}{D}_\mu=i\stackrel{\rightarrow}{\partial}_\mu+A_\mu)}
%$
%\eq
which, after 
applying the {\aut Fierz} decomposition to 2 and 3-parton correlators, 
% $$
% \label{Fierz}
% - \langle \psi(x)\, \bar \psi(z)\rangle =\frac{1}{4} \langle \bar\psi(z)\gamma_\mu \psi(x)\rangle \gamma_\mu +
% \frac{1}{4} \langle \bar\psi(z) \gamma_5 \gamma_\mu \psi(x)\rangle \gamma_\mu \gamma_5.
% $$
 %(\ref{Dirac})
 implies Equations Of Motion (EOM) 
relating the various 2 and 3-body DAs.
\vssp

\noi{\it Renormalization group equations}
\vsssp 

Back to the factorization (\ref{DAfact}) or (\ref{FactGPD_DA}) of the process in term of a DA, which symbolically reads 
\begin{equation}
     {\cal M}(Q^{2})= 
        \Phi^{*}(x,\muF) \, \otimes \, T_{H}(x,Q^{2},\muF) 
           %\,, \quad \otimes = \int_0^1 dx 
            \,, 
\label{eq:tffcf}
 \end{equation}
\nopagebreak
the arbitrariness of the factorization scale $\muF$ leads to the Efremov-Radyushkin, Brodsky-Lepage \nopagebreak equation \cite{ERBL} for  
the \alert{DA} $\Phi(u,\, \muF)$:
\[
%\begin{equation}
  \muF \frac{\partial}{\partial \muF} \Phi(x,\muF)   =
   V(x,u,\muF) \, \otimes \, \Phi(u,\muF)
         \, .
\nonumber
%\label{eq:eveq}
%\end{equation}
\]
\vsspEq

\noi{\it Collinear conformal invariance} \cite{Braun:2003rp}
\vsssp

The full conformal group $SO(4,2)$
is defined as transformations which only change the scale of the metric. In the limit 
$\alert{Q^2 \to \infty}\,,$ hadron states are replaced by a bunch of partons that are collinear to $p_1$, which thus lives along $p_2\,,$ implying that $z$ is the only remaining variable.
The transformations which map the light-ray in the $p_2$ direction into itself is the collinear subgroup of the full conformal group $SO(4,2)\,,$ that is  
 \alert{$SL(2, \R)$}, made of
translations $z \to z+c,$
dilatations $z \to \lambda \, z$ and
special conformal transformations 
$ z \to z'= z/(1+ 2 \, a \, z)\,.$ The Lie
 algebra of $SL(2, \R)$ is $O(2,1)\,.$ 
One remaining additional generator commutes with the 3 previous one: the collinear-twist operator.
Interestingly, the light-cone operators which enters the definition of  DAs
can be expressed in terms of a basis of conformal operators. Since
conformal transformations commute with exact EOM (they are not renormalized), EOM can be solved exactly
(with an expansion in terms of the conformal spin $n+2$). For example
the twist 2 DA for $\rho_L$ can be expressed, for unpredicted $a_n^\parallel(\mu)$, as \cite{OBF}
\[
%\beq
%\label{devDAconf}
\phi_\parallel(u, \,\mu_0) = 6 \,u \,\bar{u} \sum\limits^\infty_{n=0} a_n^\parallel(\mu) \, C^{3/2}_n(u- \bar{u}) \qquad C^{3/2}_n = \mbox{{\aut Gegenbauer} polynomial}\,.
%\eq
\]
%{\aut \small Ohrndorf '82; Braun, Filyanov '90} \hfill {\small but
%$a_n^\parallel(\mu)$ are unpredicted}
Since the \structure{Leading Order} renormalization of the conformal operators 
is diagonal in the conformal spin (counterterms are tree level at this accuracy and they
thus respect the conformal symetry
of the classical
theory), this implies that
\[
%\beq
%\label{solDAconf}
{\phi_\parallel(u, \,\mu) = 6 \,u \,\bar{u} \sum\limits^\infty_{n=0} a_n^\parallel(\mu_0) \left(\frac{\alpha_s(\mu)}{\alpha_s(\mu_0)}   \right)^{\gamma_n^{(0)}/\beta_0}\!\! C^{3/2}_n(u- \bar{u})\stackrel{\mu \to \infty}{\longrightarrow} 6 \,u \, \bar{u}}
\mbox{ asymptotic DA}
%\eq
\]
with the anomalous dimensions
%\centerline
\[
%\beq
%\label{anomalous}
{\gamma_n^{(0)}=C_F \left(1 -\frac{2}{(n+1)(n+2)} + 4 \sum\limits_{m=2}^{n+1}\frac{1}m    \right)} \,.
%\eq
\]
At \structure{Next to Leading Order} conformal symetry is broken; studying conformal anomalies provides the NLO anomalous dimensions and the corresponding ERBL kernels \cite{BFM}.
% \ {\aut \small Belitsky, Freund, M\"uller '99 '00} 

\section{The specific case of QCD at large $s$}

\noi{\it Theoretical
  motivations and $k_T$-factorization}
\vsssp

\begin{figure}
\psfrag{g1}[cc][cc]{$\gamma^*(q_1)$}
\psfrag{g2}[cc][cc]{$\gamma^*(q_2)$}
\psfrag{p1}[cc][cc]{$\rho(k_1)$}
\psfrag{p2}[cc][cc]{$\rho(k_2)$}
\psfrag{l1}[cc][cc]{$l_1$}
\psfrag{l1p}[cc][cc]{$-\tilde{l}_1$}
\psfrag{l2}[cc][cc]{$l_2$}
\psfrag{l2p}[cc][cc]{$-\tilde{l}_2$}
\psfrag{ai}[cc][cc]{$\stmath \beta^{\,\nearrow}$}
\psfrag{bd}[cc][cc]{$\stmath \alpha_{\,\searrow}$}
\psfrag{k}[cc][cc]{$\hspace{-.1cm} k$}
\psfrag{rmk}[lc][cc]{$\hspace{-.4cm}r-k \hspace{3cm} \int d^2 k_\perp$}
%\psfrag{oa}[cc][cc]{\raisebox{0.46
   % \totalheight}{${\tiny \alpha_k \ll \alpha_{\rm quarks}}$}}
\psfrag{oa}[cc][cc]{\raisebox{.2cm}{\footnotesize${}\quad {\tiny \stmath \alpha \ll \alpha_{\rm quarks}}$}}
\psfrag{ou}[ll][ll]{%$\slashchar{p}_1$ \quad 
\hspace{1cm} $\Rightarrow$ set $\alpha=0$ and  $\int d\beta$ }
\psfrag{ob}[cc][cc]{\raisebox{-1.5
    \totalheight}{\footnotesize$\quad {\tiny \stmath\beta \ll \beta_{\rm quarks}}$}}
 %\psfrag{ob}[cc][cc]{${} \quad {\tiny \stmath\beta_k \ll \beta_{\rm quarks}}$}
\psfrag{od}[ll][ll]{%$\slashchar{p}_2$ \quad 
\hspace{1cm} $\Rightarrow$ set $\beta=0$ and  $\int d\alpha$}
\begin{center}
\hspace{-2cm}\epsfig{file=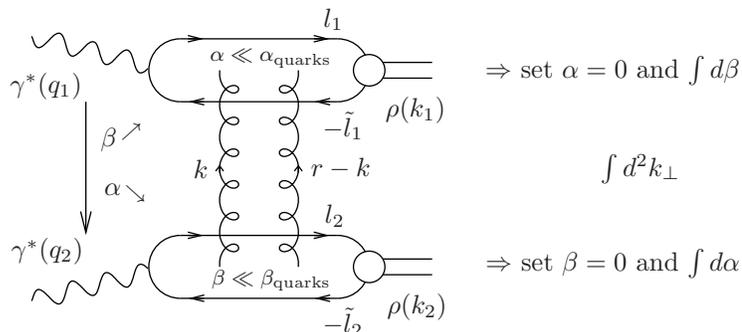,width=5.5cm}
\vspace{-.8cm}
\end{center}
\caption{$k_T$-factorization in the case of $\gamma^* \, \gamma^* \to \rho \, \rho$.}
\label{FigkT}
\end{figure}

The dynamics of
\alert{QCD} in the perturbative {\aut Regge} limit 
 is governed by gluons 
(dominance of spin 1 exchange in $t$ channel).
{\aut BFKL} Pomeron (and extensions: NLL, saturations effects, ...) is expected to dominate with respect to {\aut Born} order at
large relative rapidity in any diffractive or inclusive process.
In this regime, the key tool is the $k_T$-factorization, shown in Fig.\ref{FigkT}
in the case of 
$\gamma^* \, \gamma^* \to \rho \, \rho$.
Using the {\aut Sudakov} decomposition (\ref{Sudakov})
%\alert{$k = \alpha \,p_1 + \beta \, p_2 + k_\perp$}\  {\scriptsize ($p_1^2=p_2^2=0,\,  2 p_1 \cdot p_2=s$)}
for which ${\stmath d^4k= \frac{s}{2} \, d \alpha \, d\beta \, d^2k_\perp}$ 
and noting that $t-$channel gluons with \alert{non-sense} polarizations  ($\epsilon_{\alert{NS}}^{up / down}=\frac{2}s \, p_{2\, /1}$) dominate \alert{at large $s$}, and then rearanging the $k$-integration as shown in Fig.\ref{FigkT}
%and rearrange integrations \alert{in the large $s$ limit}: \ \ 
%At large $s,$  $r = r_\perp$ and
leads to the impact representation 
% \vspace{-.45cm}
% \hspace{7.5cm}
% {\scriptsize  $\kb$ = Eucl. $\leftrightarrow $ $k_\perp$ = Mink.}
% \vssp
\beq
\label{impactRep}
{\cal M} = is\;\int\;\frac{\stmath{d^2\,\kb}}{(2\pi)^2\stmath{\kb^2\,(\rb -\kb)^2}}
\alert{{\Phi}^{\gamma^*(q_1) \to \rho(p^\rho_1)}}(\stmath{\kb,\rb -\kb})\;
\alert{{\Phi}^{\gamma^*(q_2) \to \rho(p^\rho_2)}}(\stmath{-\kb,-\rb +\kb})
\eq
where
$\alert{{\Phi}^{\gamma^*(q) \to \rho(p^\rho)}}$ is the   \ $\gamma^*_{L,T}(q) \, g(k_1) \to \rho_{L,T}(p^\rho)\, g(k_2)$ \alert{impact factor} (with $k_\perp^2=- \kb^2$). 
\alert{QCD gauge invariance} implies, for colorless probes,  
that the \structure{impact factor should} \alert{vanish} when \alert{$\kb \to 0$} or \alert{$\rb-\kb \to 0$}.
In particular, at twist 3 level (for $\gamma^*_T \to \rho_T$ transition), gauge invariance
is a non-trivial statement which \alert{requires 2- {\it and} 3-parton correlators}.
Recently, HERA provided data
 for vector mesons with detailled \alert{polarization} studies, in particular for our example $\gamma^*_{L,T}+p\to \rho_{L,T}+p$ \cite{expVectorHigh}.
It exhibits a total cross-section which strongly 
\alert{decreases with $Q^2$}, with a
dramatic increase with $W^2=s_{\gamma^*P}$.
 The transition from soft to hard regime 
is
governed by $Q^2$. 
 The transitions
$\gamma_L^* \to \rho_L$,   
$\gamma_{T(-)}^* \to \rho_{T(-)}$ and $\gamma_{T(+)}^* \to \rho_{T(+)}$ dominate with respect to any other possible transition, as expected from SCHC
discussed above. In particular
at $t=t_{min}$ one can experimentally distinguish two transitions:
$
\gamma_L^* \to \rho_L$ which dominate (\alert{twist 2 dominance}) and the 
$\gamma_{T(\pm)}^* \to \rho_T(\pm)$ which is sizable, although of twist 3. This calls for detailled studies beyond the applicability of the collinear factorization theorem.
\vssp

\noi{\it Phenomenological applications: meson production at  HERA}
\vsssp

The production of mesons in diffraction-type experiments
at
{\aut HERA} has been studied extensively in various situations \cite{expVectorHigh, rho0HERMES}.
In the safe case, like $J/\Psi$ photoproduction,  collinear factorization holds 
%($u \sim 1/2:$ non-relativistic limit for a heavy bound state) 
and 
 combined with $k_T$-factorization  a consistent description of H1 and ZEUS data was obtained \cite{JPsiTheory, large_t_Light_or_HeavyMeson}.
%%in the large $|t|$ limit (which plays the role of a hard scale).
% {\aut \small Ryskin '93; Frankfurt, Koepf, Strikman '98; Ivanov, Kirschner, Sch\"afer, Szymanowski '00; Motyka, Enberg, Poludniowski '02}
In the more intricate case of
exclusive light vector meson ($\rho$, $\phi$) photoproduction at \alert{large $t$},  
relying on 
 $k_T$-factorization, one can describe
%{%\footnotesize 
%\aut \small Forshaw, Ryskin '95; Bartels, Forshaw, Lotter,  W\"usthoff '96; Forshaw, Motyka, Enberg, Poludniowski '03} 
H1 and ZEUS data, which  
%\item test of {\aut BFKL} at \alert{large $t$} (=hard scale)
 seem to favor {\aut BFKL} \cite{large_t_Light_or_HeavyMeson, large_t}. 
	 One needs to regularize end-point singularities for $\rho_T$,  using for example a quark mass $m=m_\rho/2\,,$ and a
rather poor understanding of the whole spin density matrix has been achieved.
The exclusive vector meson electroproduction
${\stmath \gamma^*_{L,T}+p\to \rho_{L,T}+p}$
has been described \cite{GK}
%only studies with every $\rho$ polarization: 
%{\aut \small Goloskokov, Kroll '05} 
based on the ICA for DA coupling and 
collinear factorization with GPDs, as explained above, without any use of $k_T$ factorization. However, it turns out that at moderate value of $s$, HERMES \cite{rho0HERMES}
measured the interference phase between $L \to L$ and  $T \to T$ transitions %interference between $L \to L$ and  $T \to T$ exhibit a real and and imaginary part 
which cannot be described within perturbative QCD at the moment.

A full twist 3 treatment of $\rho$-electroproduction in $k_T$-factorisation is possible
\cite{usProceedingsANDShort}.
It relies on the computation of the 
 $\gamma^*_T \to \rho_T$ impact factor at twist 3 
%{\aut \small Anikin, Ivanov, Pire, Szymanowski, S.W. to appear}
including  consistently  all twist 3 contributions, i.e. 2-parton and 3-parton correlators. This gives a gauge invariant impact factor, and an amplitude which is free of end-point singularities due to the presence of $k_T$.
\vssp

\noi{\it Exclusive processes at Tevatron, RHIC, LHC, ILC}
\vsssp

Exclusive $\gamma^{(*)} \gamma^{(*)}$ processes are golden places for testing QCD at large $s$, in particular at Tevatron, RHIC, LHC and ILC.
Several 
proposals in order to test perturbative QCD in the large $s$ limit
($t$-structure of the hard Pomeron, saturation, Odderon...) have been made, including
${\stmath \gamma^{(*)}(q)+\gamma^{(*)}(q^\prime)\to  J/\Psi \, J/\Psi}$ \cite{KM},
%
%{\aut \small Kwiecinski, Motyka '98}
or
${\stmath \gamma^*_{L,T}(q)+\gamma^*_{L,T}(q^\prime)\to \rho_L(p_1)+\rho_L(p_2)}$ process in $e^+ \, e^-\to e^+ \, e^- \rho_L(p_1)+\rho_L(p_2)$ with double tagged lepton at {\aut ILC} \cite{IP, PSW}.
%{\aut \small Pire, Szymanowski, S. W. '04; Pire, Szymanowski, Enberg, S. W. '06; Ivanov, Papa '06;  Segond, Szymanowski, S. W. '07}
This could be
feasible at {\aut ILC} (high energy and high luminosity), with an expected {\aut BFKL NLL} enhancement with respect to {\aut Born} and {\aut DGLAP}.
The elusive \alert{Odderon}  ($C$-parity of Odderon = -1) is hard to reveal directly
when entering in the amplitude of a process. When considering processes where it 
enters linearly, through interference with the Pomeron, the signal becomes more favorable \cite{odderon}, as in  ${\stmath \gamma+\gamma\to \pi^+ \pi^- \pi^+ \pi^-}$: a
$\pi^+ \pi^-$ pair has no fixed $C$-parity, allowing for both Odderon and Pomeron exchange, which can interfere  in the charge asymmetry \cite{Pire:2008xe}.
%
%{\aut \small Pire, Schwennsen, Szymanowski, S. W. '07}
%
%see the talk of {\aut F. Schwennsen}
More generally exclusive ultraperipheral processes are very promising.
%see the talk of {\aut J. Nystrand}

\section{Light-Cone Collinear Factorization}

%\subsection{Light-Cone Collinear Factorization}
\begin{figure}[h]
\psfrag{rho}[cc][cc]{$\rho$}
\psfrag{k}[cc][cc]{}
\psfrag{rmk}[cc][cc]{}
\psfrag{l}[cc][cc]{$\ell$}
\psfrag{q}[cc][cc]{}
%\psfrag{q}[cc][cc]{$q$}
\psfrag{lm}[cc][cc]{}
\psfrag{H}[cc][cc]{$ H_{q \bar{q}}$}
\psfrag{S}[cc][cc]{$ \Phi_{q \bar{q}}$}
\scalebox{.8}{\hspace{0cm}\begin{tabular}{ccccc}
\raisebox{0cm}{\epsfig{file=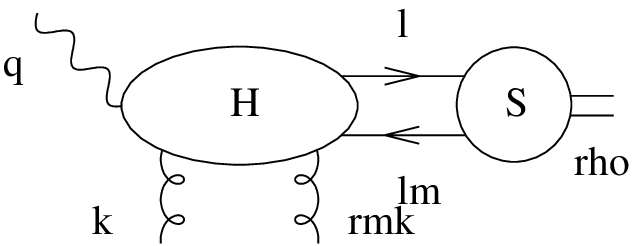,width=4.5cm}}
&  \hspace{-.2cm}\raisebox{.9cm}{$\longrightarrow $} \
&
\psfrag{lm}[cc][cc]{\raisebox{.2cm}{$\quad \,\,\, \, \, \Gamma \ \,\, \Gamma$}}
\hspace{-.6cm}\epsfig{file=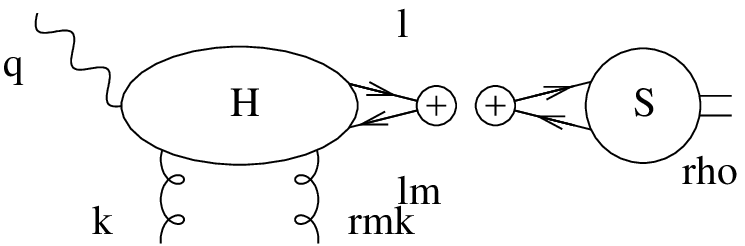,width=5.5cm}
&\hspace{-.2cm}\raisebox{.9cm}{$+$} 
&
\psfrag{H}[cc][cc]{{$H^\perp_{q \bar{q}}$}}
\psfrag{S}[cc][cc]{\scalebox{1}{$\Phi^\perp_{ q \bar{q}}$}}
\psfrag{lm}[cc][cc]{\raisebox{.2cm}{$\quad \,\,\, \, \, \Gamma \ \,\, \Gamma$}}
\hspace{-.5cm}
\epsfig{file=wallon_samuel.FiertzHSqq_rhofact.eps,width=5.5cm}
\end{tabular}
%\caption{Factorization of 2-body contributions in the  example of the $\gamma^* \to \rho$ impact factor.}
%\label{fig:Factorized2body}
}
\caption{Factorization of 2-parton contributions in the  example of the $\gamma^* \to \rho$ impact factor.}
\label{Fig:Factorized2body}
\end{figure}
There are basically two ways of dealing with collinear factorization when including higher twist corrections. 
  The \alert{Light-Cone Collinear Factorization} developped first for polarized DIS \cite{EFP},  is self-consistent for exclusive processes \cite{Anikin:2000em, Anikin:2001ge}, while non-covariant, and very efficient for practical computations \cite{usProceedingsANDShort}.
%{\aut \small Anikin, Ivanov, Pire, Szymanowski, S.W. '09}  
%{\aut Ellis, Furmanski, Petronzio '83; Efremov, Teryaev '84}
 Using the {\aut Sudakov} decomposition  ($p=p_1, \,n=2 \, p_2/s$ thus $p \cdot n =1$)
\beq
\label{SudakovPN}
\begin{array}{cccccccc}
\ell_\mu &=&\alert{u \,p_\mu}  &+& {\twist3 \ell^\perp_\mu} &+& (\ell\cdot p)\, n_\mu ,&
\quad \alert{u}=\ell\cdot n \\
\\
%{\small 
\text{\structure{scaling:}} & & {\stmath ~1} & & { \stmath ~1/Q} & & {\stmath ~1/Q^2}
%}
\end{array}
\eq
one decomposes $H(k)$ around the \alert{$p$} direction:
\beas
% H(l) = {\alert{H(x p)}} &+& {\twist3 \frac{\partial H(l)}{\partial l_\alpha} \biggl|_{l=x p}\biggr. \, (l-x \,p)_\alpha} + \ldots
% \quad \text{with} \,\,\,\, (l-x\, p)_\alpha \approx{\twist3 l^\perp_\alpha} \\
% \text{\small \alert{twist 2}}&&\text{\small {\ktwist3 kinematical twist 3} {\it and} \ \structure{genuine twist 3}}
H(\ell) = H(u \, p) &+& \frac{\partial H(\ell)}{\partial \ell_\alpha} \biggl|_{\ell=u p}\biggr. \, (\ell-u \,p)_\alpha + \ldots
 \quad \text{with} \,\,\,\, (\ell-u\, p)_\alpha \approx{\twist3 \ell^\perp_\alpha} 
\eeas
from which  the {\twist3  twist 3} term {\twist3 $l^\perp_\alpha$} turns after Fourier transform into the derivative of the \structure{soft term} 
%$\Rightarrow$ one will deal with %$\stmath S_{q \bar{q}}(l) +  
$\int d^4z \ e^{- i \ell\cdot z }
\langle \rho(p)|
\psi(0) \,{\twist3 i \, \stackrel{\longleftrightarrow}
{\partial_{\alpha^\perp}}} \bar\psi(z)| 0 \rangle \,.$
Using the {\aut Fierz} transformation, this gives finally a factorized expression up to
twist 3, as illustrated in Fig.\ref{Fig:Factorized2body} for 2-parton contributions in the example of the $\gamma^*\,g \to \rho \, g$ impact factor.
This requires the parametrization of 
	 matrix elements of non-local correlators  defined along the \structure{light-like prefered direction} $z = \lambda \, \alert{n}$ conjugated to  $p$. In the case of the $\rho$-electroproduction, 
 \structure{7 DAs at twist 3} (2- and 3-parton DAs) are needed.
%\item The computation of the hard part is much simpler than within the covariant approach
Their number is  reduced to a minimal set of 3 DAs when combining the 2 equations of motions
and the \alert{$n-$independency condition} \cite{EFP, Anikin:2000em, Anikin:2001ge, Anikin:2002wg} of the full factorized amplitude (which provides 2 process-independent equations).
A second approach, the Covariant Collinear Factorization \cite{BB}, 
%({\aut Braun, Ball}), 
fully covariant but less convenient when practically computing
coefficient functions, can equivalently be used. 
The dictionary and equivalence between the two approaches
has recently been obtained, 
and   explicitly checked for the $\gamma^*_T \to \rho_T$ impact factor at twist 3 \cite{usProceedingsANDShort}. 
%{\aut \small Anikin, Ivanov, Pire, Szymanowski, S.W. to appear}

\section{Conclusion}

Since a decade, there has been much progress in the understanding of \alert{hard} exclusive processes:  
%starting from forward matrix elements
 	%{\small 
 at moderate energies, combined with GPDs, \structure{there is now a framework starting from first principles to describe a huge number of processes}; 
%(for pure QCD processes as well
% as for processes where QCD effects are essential, like semileptonic decays: ex.: $B \to \pi \ell \nu_\ell$)
at high energy, \structure{the impact representation} is a powerful tool for describing exclusive processes in diffractive experiments, which are and will be essential for studying QCD in the hard Regge limit (Pomeron, Odderon, 
saturation...). 
Still, \structure{some problems remain:}
\structure{proofs of factorization have been optained only for a very few 
processes} 
(ex.: $\gamma^* \, p \to \gamma \, p\,$, $\gamma^*_L \, p \to \rho_L \, p\, , \, \gamma^* p \to J/\Psi \, p$).
For some other processes factorization is highly plausible, but not fully demonstrated at any order (ex.: processes involving Transition Distribution Amplitudes \cite{TDA}) while some processes explicitly show signs of breaking of 
factorization
(ex.:  $\gamma^*_T p \to \rho_T p$ which has end-point singularities at Leading Order).
Models and results from the lattice for the non-perturbative correlators entering GPDs, DAs, GDAs, TDAs
 are needed, even at a qualitative level.
The effect of QCD evolution and   renormalization/factorization scale, as well as studies at the full NLL order, might be relevant with the increasing precision of data in the near future. Finally, let us insist on the fact that links between theoretical and experimental communities are very fruitful, in particular at HERA, Jlab, Compass. It is now time to use the potential of high luminosity $e^+ e^-$ machine like {\aut BaBar, BELLE, BEPC-II} which \alert{are golden places for hard exclusive processes studies} in $\gamma^{*} \gamma^{(*)}$ channels.

% ****************************************************************************
% BIBLIOGRAPHY AREA
% ****************************************************************************

\begin{footnotesize}
% IF YOU DO NOT USE BIBTEX, USE THE FOLLOWING SAMPLE SCHEME FOR THE REFERENCES
% ----------------------------------------------------------------------------

\end{footnotesize}

% ****************************************************************************
% END OF BIBLIOGRAPHY AREA
% ****************************************************************************

%%%%%%%%%%%%%%%

\end{document}